\def\ga{\mathrel{\mathchoice {\vcenter{\offinterlineskip\halign{\hfil
$\displaystyle##$\hfil\cr>\cr\sim\cr}}}
{\vcenter{\offinterlineskip\halign{\hfil$\textstyle##$\hfil\cr
>\cr\sim\cr}}}
{\vcenter{\offinterlineskip\halign{\hfil$\scriptstyle##$\hfil\cr
>\cr\sim\cr}}}
{\vcenter{\offinterlineskip\halign{\hfil$\scriptscriptstyle##$\hfil\cr
>\cr\sim\cr}}}}}

\documentclass[final]{aipproc}
\layoutstyle{6x9}
\usepackage{graphicx}  

\begin{document}

\title{The CTA Observatory}

\classification{95.85.Pw; 98.70.Sa}
\keywords      {Gamma rays; Cosmic rays}

\author{J\"urgen Kn\"odlseder, on behalf of the CTA consortium}{
  address={CESR, CNRS/UPS, - 9, avenue du Colonel Roche, 31028 Toulouse, France}
}

\begin{abstract}
Ground-based gamma-ray astronomy has experienced a major breakthrough in the last
decade thanks to the advent of new generation instruments such as H.E.S.S., MAGIC, 
Milagro and VERITAS. 
A large variety of cosmic particle accelerators has been unveiled, comprising supermassive 
black holes in the centres of active galaxies, nearby star forming galaxies, 
Galactic supernova remnants and pulsar wind nebulae, and stellar binary systems housing
a compact object. 
While current instruments revealed the tips of the non-thermal icebergs in our Universe, 
a factor of 10 increase in sensitivity, improved angular resolution and an extended 
energy coverage is required to fully explore and understand the physics of cosmic 
particle acceleration. 
The Cherenkov Telescope Array (CTA) will provide these performances, by deploying two 
arrays of Cherenkov telescopes in the northern and southern hemispheres, allowing
full-sky coverage.
In this paper we summarize the project status and present the science prospects of the CTA 
observatory. 
\end{abstract}

\maketitle

\section{Cosmic Particle Accelerators}

Our Universe is full of particle accelerators.
Cosmic particle acceleration is witnessed most directly by the presence of cosmic-ray (CR)
particles near the Earth, which are readily observed by direct measurements above
the Earth's atmosphere \cite{alcaraz00,wang02},
or by observation from the ground of their induced particle cascades in the Earth's atmosphere
\cite{becker09}.
While those measurements provide valuable information about the CR energy spectrum and
composition, they offer limited insight into the nature of the CR sources as the
particles get considerably deflected by magnetic fields on their ways from the acceleration
site to the Earth.
An observable manifestation of this deflection is the radio synchrotron emission 
that arises when relativistic electrons spiral through magnetic fields.
Such radiation is seen from a large variety of objects, covering 
the Sun, 
nova and supernova explosions,
gamma-ray bursts (GRBs),
supernova remnants (SNRs),
pulsar wind nebulae (PWNe),
relativistic stellar binary systems,
galaxies (and in particular our own Galaxy from which widespread diffuse emission is observed),
active galactic nuclei (AGN), and
galaxy clusters.
The ubiquity of synchrotron emission is a clear demonstration that particle acceleration is 
omnipresent in the Universe.

Nevertheless, we still ignore where and how the large majority of CRs are accelerated 
\cite{halzen10}.
Furthermore, it remains a big mystery how nature manages to concentrate up to 
$\sim10^{20}$ eV of kinetic energy into a single particle that is observed in CRs at 
Earth \cite{stanev07}.
We also have still a poor understanding of how CRs propagate through the Universe
\cite{kotera08} or within our Galaxy \cite{cowsik10}, and about how they impact their 
environment \cite{indriolo09,papadopoulos10,sijacki08}.
Our ignorance is partly related to the fact that synchrotron emission traces generally 
only the leptonic component of CRs.
Leptons, however, constitute only about 1\% of the CR particles while the remaining
99\%, mainly protons and helium nuclei, escape observations.
Energetically, however, only the hadronic component is relevant.

Fortunately there are gamma rays.
First of all, gamma rays (photon energies $\ga511$~keV) provide a cleaner signal of 
particle acceleration than do the radio signals due to the absence of competing thermal 
emission processes at these high energies.
Even more important, gamma rays give access to the hadronic CR component which is 
traced by a (Doppler broadened) gamma-ray line centred at $68$~MeV (in photon flux)
that arises from the decay of $\pi^0$ mesons generated in inelastic collisions of CR hadrons
with interstellar matter and radiation.
Competition arises in the gamma-ray domain, however, from 
non-thermal Bremsstrahlung,
inverse Compton scattering or
curvature radiation of relativistic electrons interacting with
gas,
photon fields
or strong magnetic fields, respectively.
The drawback of this competition is that hadronic and leptonic signatures are blended.
Disentangling the various emission processes thus requires the broadest possible 
energy coverage in the gamma-ray domain, eventually aided by lower energy 
observations (radio, optical, infrared, X rays), to allow the spectral separation of the
components.
Having several different emission processes is, however, also an advantage: it
allows for a large variety of particle accelerators to be studied, and gives access to
information about the physical conditions of the local environment.

\section{Existing Facilities}

The study of cosmic particle accelerators is thus by nature a multiwavelength endeavour.
Only recently, a complete spectral coverage of the gamma-ray domain has become
available, thanks to the advent of new generation ground-based gamma-ray
telescopes such as H.E.S.S. \cite{aharonian06}, MAGIC \cite{baixeras04}, 
Milagro \cite{atkins04} and VERITAS \cite{weekes02}.
These instruments cover the energy range between a few tens of GeV up to $\sim10$~TeV
-- the Very High Energy (VHE) domain -- and rely on the detection of electromagnetic showers
created by interactions of VHE gamma rays with the Earth's atmosphere.
So far\footnote{
  See http://tevcat.uchicago.edu/ for an up-to-date list of VHE sources.},
more than 100 cosmic particle accelerators have been unveiled by this technique, 
comprising Galactic (PWN, SNR, gamma-ray binaries) and extragalactic objects (starburst 
galaxies and AGN).
This demonstrates that current VHE telescopes have passed the critical sensitivity 
threshold that enables the study of VHE particle accelerators in the Universe.

The {\it Fermi} Gamma-Ray Space Telescope, launched in 2008 and operating since then
at lower energies between 30~MeV to 300~GeV \cite{atwood09}, 
indicates that this is only the tip of the iceberg.
The {\em Fermi} Large Area Telescope first source catalogue \cite{abdo10a} lists 1451
sources, comprising large populations of pulsars and AGN in addition to PWN, SNR,
gamma-ray binaries, globular clusters, normal galaxies, starburst galaxies and radio
galaxies.
A comparable number of sources is also expected in the VHE domain, provided that the
detection sensitivity can be improved by a factor of $\sim10$.
With such an improvement, the required broadband energy coverage would become
available for a large number of sources, enabling a thorough study of the physics of 
cosmic particle accelerators in the Universe.

\section{The CTA Observatory}

The Cherenkov Telescope Array (CTA) will provide the required increase in sensitivity, 
together with improvements in angular resolution and an extension of the energy range.
CTA will consist of two arrays of Cherenkov telescopes located at sites in the northern and 
southern hemispheres, allowing full-sky coverage.
The increase in sensitivity will be achieved by the deployment of large numbers (50 to 100)
of Cherenkov telescopes, while the extension of the energy range will be accomplished
by using telescopes of different sizes.
An increase of angular resolution with respect to existing facilities will be achieved by
improved imaging of the air shower, both in terms of resolution and photon statistics.
For a detailed description of the CTA design and the expected performance, the reader should 
refer to \cite{hofmann10}. 
 
CTA will be operated as a proposal-driven open observatory, with a Science Data Centre
providing transparent access to data, analysis tools and user training.
A deep survey of the Galactic plane, and possibly, a more shallow all-sky survey will
complement these observations, possibly in form of a core or key program.
The CTA project recently entered the preparatory phase, funded by the European Commission
under the Seventh Framework Programme (FP7), which will address a number of crucial
prerequisites for the approval, construction and operation of CTA.
This phase will last for 3 years and will deliver a complete and detailed implementation
plan for the CTA infrastructure.
Array deployment may then start from 2014 on, provided that funding is secured.

\section{Science prospects}

The main science themes that will be addressed by CTA are summarized in the following 
sections.
For a more comprehensive discussion of the CTA science case, the reader should refer to
\cite{hofmann10}.

\subsection{Origin of cosmic rays and their role in the Universe}

The standard paradigm, mainly based on energetic grounds, is that Galactic CRs are
accelerated in the shocks generated by supernova explosions \cite{ginzburg64}.
Gamma-ray emission is indeed detected from a growing number of Galactic
SNRs, yet the nature of the underlying emission process (hadronic or leptonic) is still 
under debate \cite{zirakashvili10}.
Moreover, Galactic CRs reach energies of at least several $10^{15}$ eV (PeV), 
which in turn should produce VHE gamma rays with a relatively flat energy
spectrum extending to hundreds of TeV.
Sources that show such characteristics have been dubbed PeVatrons.
So far, however, not a single SNR PeVatron has been detected.
This may still be commensurate with the estimate that only $\sim10$ such 
objects exist today in our Galaxy, since rapid escape of PeV particles
limits their acceleration lifetime in SNR shocks to only several hundred years.
The deep CTA survey of the Galactic plane should unveil these $\sim10$
SNR PeVatrons, provided that the standard paradigm for the Galactic CR origin 
is correct.

Several PeVatron candidates have indeed already been detected by the existing VHE 
facilities, but they all seem to be associated with PWNe \cite{camilo09,abdo10b}.
Also the Crab, which is considered to be the only known Galactic PeVatron 
\cite{aharonian04}, is a PWN.
Conventionally, VHE emission of PWNe is interpreted to be predominantly of leptonic
origin \cite{baring10} although there are arguments suggesting the presence of
substantial amounts of relativistic protons in PWNe \cite{hoshino92}.
The observed anisotropy in the arrival directions of CRs with energies $\sim20$ TeV
\cite{desiati10} may be a hint for nearby Galactic CR sources, for which the Vela and 
Geminga PWNe appear to be plausible candidates \cite{halzen10}.
The detection of a clear hadronic signal by CTA from PWNe would thus be an 
invaluable piece of evidence to establish their role as Galactic CR sources.

PWNe constitute also the biggest class of identified Galactic VHE sources, and CTA
is expected to detect on the order of $\sim100$ objects, allowing for the first time
comprehensive population studies.
A comparably sized sample of Galactic SNRs should be visible to CTA, enabling studies
of particle acceleration as function of SNR age and environment.
For example, if our general picture of SNR evolution is correct, the position of the cutoff
in the VHE spectrum should depend on the age of the SNR and on the magnetic field at the 
shock.
A SNR population study will allow testing this hypothesis and constraints to be placed
on the physical parameters of SNRs, in particular on the magnetic field strengths.

CTA offers also the possibility to directly observe the diffusion of CRs in the Galaxy.
While travelling from the accelerator to the target, the spectrum of CRs is a strong
function of time, distance to the source, and the local diffusion coefficient.
Depending on the values of these parameters, varying gamma-ray spectra are expected
from the environment surrounding CR accelerators, in particular from massive molecular clouds
that provide thick targets for CR hadronic interactions \cite{gabici09}.
As a proof of principle, surprisingly small CR diffusion coefficients have been recently 
inferred from observations of gamma-ray emission from molecular clouds near two 
Galactic SNRs \cite{gabici10}.
CTA will enable detailed mappings of VHE emission around potential CR accelerators, 
enabling thus an experimental determination of the local diffusion coefficients and/or the 
local CR injection spectra in our Galaxy.

At a larger scale, CR diffusion can also be studied by CTA in nearby galaxies of the Local
Group, such as the Magellanic Clouds or M31.
Recent {\em Fermi} observations of the Large Magellanic Cloud revealed
a surprisingly close confinement of CRs around star forming regions, which also may
point towards a small proton diffusion length \cite{abdo10c}.
With its superior angular resolution, CTA will be able to refine these studies by providing
a detailed mapping of the CR density in these nearby galaxies.
CTA will also enable comparative studies of external galaxies, such as the Local Group
galaxies and nearby starburst galaxies.
These comparative studies will provide important clues about the key parameters of CR 
acceleration and transport \cite{lacki10}.
They also will elucidate the role of CRs in galactic feedback processes 
\cite{papadopoulos10,socrates08}.

\subsection{Nature and variety of particle acceleration around black holes}

Accretion of matter onto a black hole provides one of the most efficient energy source
known in the Universe.
The most massive black holes in our Universe are presumably hosted in the centres
of active galactic nuclei (AGN), and accreting onto these black holes readily explains
the tremendous luminosity of these objects \cite{dermer09}.
AGN often show jets of relativistic plasma, and jet sources are often most luminous
at gamma-ray energies.
AGN with jets that are aligned with the line-of-sight to within a few degrees are dubbed
blazars, and those present the dominant class of extragalactic VHE emitters known so
far.
The observed fast variability of the VHE flux from blazars, down to minute time scales 
\cite{aharonian07}, indicates that gamma-ray production occurs in spatially
confined region, with sizes as small as a few times the Schwarzschild radius of
the black hole.
What remains unknown is how the relativistic jets are launched, what their structure
and composition is, and by what physical mechanism the particles are accelerated to
very high energies \cite{hardee10}.
Multi-wavelength observations with high temporal and spectral resolution can help to
answer these questions.
By giving access to the VHE emission for a large population of blazars, CTA will provide 
new insights into the physics that is driving these sources.
In particular, CTA will be able to probe variability time scales well below minutes, putting 
constraints on acceleration and cooling times, instability growth rates, and the time evolution 
of shocks and turbulences.

As VHE blazars are seen to considerable distances, their gamma rays may also
be used to study the diffuse ultraviolet to infrared radiation along their lines-of-sight.
VHE gamma rays travelling from remote sources interact with photons of the
extragalactic background light (EBL) via e$^+$e$^-$ pair production, leading to an 
energy-dependent absorption of the intrinsic source spectrum.
If this intrinsic spectrum is known, the observations may provide a measure
about the integrated EBL density towards the source \cite{raue10}.
First attempts to constrain the EBL density from VHE observations are promising
\cite{aharonian06a}.
CTA will characterise a sufficiently large sample of extragalactic VHE sources that will
constrain the EBL density as function of redshift, complementing thus direct measurements
in the infrared domain that are hampered by strong Galactic and zodiacal
foreground emissions.

More locally, radio galaxies have recently emerged as a new class of VHE emitting AGN
\cite{aharonian06b}.
Jets of radio galaxies show only small -- if at all -- relativistic boosting as they make
larger angles to the line-of-sight than the jets of blazars.
This considerably reduces the jet luminosity, and thus only nearby radio galaxies are 
detectable in gamma rays; on the other hand, most AGN in the local Universe 
are radio galaxies, hence the number of potential sources is large.
Given the proximity and the observing geometry of the sources, detailed spatially resolved
studies of individual jet components become possible in gamma rays.
Furthermore, as the relativistic boosting is no longer a poorly constrained parameter of 
the problem, more stringent constraints on the VHE emission physics and geometry can
be derived.
CTA will allow to resolve the outer and inner kpc jet structure of nearby radio galaxies, 
enabling to spatially pin down the site of the emission.
With the help of simultaneous multi-wavelength observations and (temporal) correlation
studies, different sections of the jet and the core can be probed, down to the smallest 
sub-pc (milli-arcsecond) scale, only accessible to VLBA radio observations.
As proof of principle, a multi-wavelength observing campaign of the radio galaxy M87 
combining VHE, X-ray and VLBI radio observations has recently localised the observed
VHE emission to within a few Schwarzschild radii of the black hole \cite{acciari09}.

Relativistic jets are also the plausible origin of the gamma-ray emission that has
recently been detected from the Galactic microquasar Cyg X-3 \cite{abdo09a}.
Furthermore, a handful of VHE gamma-ray emitters are known to be Galactic binary
systems, consisting of a compact object (neutron star or black hole) orbiting a massive
star.
Whilst jet emission powered by accretion onto the compact object has been suggested
to explain the observations \cite{dermer06}, there are arguments that favour the
VHE emission to originate in the interaction of relativistic outflows from highly magnetised 
neutron stars (a.k.a. pulsar wind) colliding with the wind of the massive star \cite{dubus06}.
High-mass X-ray binary population studies predict that CTA may detect a dozen new
systems of that kind \cite{cerutti09}.
But already the observations of the few known systems with improved sensitivity will
provide tight constraints on the spectral dependency of their orbital modulations, offering
deeper insights into the underlying physics \cite{dubus10}.
Furthermore, the CTA deep Galactic plane survey has the potential to
reveal more sources, and continuous monitoring of key objects (such as Cyg X-3 or Cyg X-1)
may allow to catch also flaring VHE emission that, in combination with multi-wavelength 
observations, will inform about the link between accretion and particle acceleration around
compact objects.

\subsection{Physics beyond the horizon}

It is difficult to speculate about the unknown, and definitely we cannot accurately predict
how much CTA will unveil about any physics that is beyond our ``standard model'' of the 
world. 
Gamma rays, however, hold the potential to reveal properties of the elementary
particles that make up our Universe because photonic signatures of particle interactions, decays 
and annihilations show up in this energy range.
Weakly interacting massive particles, the leading category of particle dark matter
candidates, may give rise to continuum and/or line signatures in the VHE domain that are
in principle observable by CTA \cite{bringmann09}.
An alternative dark-matter candidate are axions which are expected to oscillate
into photons (and viceversa) in the presence of magnetic fields \cite{dicus78}.
These oscillations could distort the spectra of gamma-ray sources, providing an
indirect method to detect these particles from observations with CTA \cite{sanchez09}.

Another domain of fundamental physics for which CTA may provide important constraints
is the validation of Lorentz invariance, a principle on which the theory of special
relativity is based \cite{mattingly05}.
Some theories of quantum gravity predict violation of Lorentz invariance which would
manifest as an energy dependence of the speed of light.
While the most stringent constraints on this energy dependence come today from arrival 
time delays measured by {\em Fermi} in GRBs \cite{abdo09b},
CTA may provide potentially more stringent constraints from precise timing of AGN flares.

\section{Conclusions}

The CTA observatory is the logical next step in the exploration of the high-energy
Universe, and will promote VHE observations to a public tool for modern astronomy. 
CTA will explore the VHE domain from several tens of GeV up to more than
10 TeV with unprecedented sensitivity and angular resolution, enabling a 
comprehensive understanding of cosmic particle acceleration physics at
various scales, distances and time scales.
Major advances are expected in understanding the origin of Galactic cosmic rays, 
their propagation within galaxies, and their impact on their environment.
Particle acceleration in the vicinity of black holes will be explored in a large variety of
sources, and interactions and feedback effects of the particles on their surroundings 
will be explored.
CTA will also probe physics beyond the established horizon, holding promises for
a better understanding of the ultimate laws that govern the Universe.
The CTA project just started the 3-years lasting preparatory phase, and array
deployment could begin as early as 2014, with a full observatory operational
before the end of this decade. 
Early science may be optimistically expected from 2015 on.

\begin{theacknowledgments}
I particularly thank G. Dubus, B. Giebels and B. Khelifi for a careful reading of the manuscript.
In addition, I want to thank all my colleagues from the CTA consortium for the tremendous 
work being done during the Design Study and the now commencing Prepatory Phase. 
The support of the involved National funding agencies and of the European community 
is gratefully acknowledged, as is the support by the H.E.S.S. and MAGIC collaborations 
and the interested parties from the US.
\end{theacknowledgments}


\end{document}